\documentstyle[aps,twocolumn,epsf]{revtex}

\def\uu{^{\mbox{ }}}
\def\s{\sigma}
\def\ss{\Sigma}
\def\up{\uparrow}
\def\dd{\downarrow}
\def\las{\langle}
\def\ras{\rangle}

\def\si{\mbox{sin}}
\def\co{\mbox{cos}}
\def\p{^{\prime}}

\begin{document}
\title{Spin-polarized tunneling through a thin-film}
\author{Alireza Saffarzadeh\thanks{E-mail: a-saffarzadeh@cc.sbu.ac.ir}}
\address{Department of Physics, Shahid Beheshti University, 19839, Tehran, Iran}
\date{12 June 2000}
\maketitle

\begin{abstract}
The effect of spin-disorder scattering on perpendicular transport in a magnetic
monolayer is considered within the single-site Coherent Potential Approximation
(CPA). The exchange interaction between a conduction electron and localized
moment of the magnetic ion is treated with the use of the {\it s-f} model.
Electron-spin polarization is evaluated in the tunnel current which comes from
the different densities of spin-up, spin-down conduction electrons at the Fermi
level in a ferromagnetic semiconductor (EuS). Calculated results are compared
with some tunneling experiments.
\end{abstract}
{\it Keywords}: Electron-spin polarization; Spin-filter effect; Spin-disorder
scattering

\section{Introduction}
There has been renewed interest in spin-polarized transport over the last
decade. This interest comes in part because of a wide range of novel phenomena,
e.g., the giant and colossal magnetoresistance \cite{Baibich}, spin-polarized
tunneling experiments \cite{Julliere}, and exchange coupling (or spin currents)
\cite{Gr}. One of the most fundamental properties of
spin-polarized transport in a ferromagnet is the polarization in the density of
states at the Fermi energy.
This polarization enters either directly or
indirectly into most transport calculations. In particular, since tunneling
experiments measure the density of states, they should provide a direct measure
of this polarization.
In the case of ferromagnet-insulator-ferromagnet tunneling
experiments one measures the product of the spin polarizations. However, in
ferromagnet-insulator-superconductor tunneling experiments where the density of
states in the superconductor is Zeeman split by a field in the plane of the
film, one can in principle measure directly the spin polarization in the
density of states. In these experiments the spin polarization was
attributed to the difference in the spin densities of states of the itinerant
electrons in the ferromagnets at the Fermi energy.
In contrast to these experiments there have been a series of experiments where
electron-spin polarization (ESP) of the tunneling current has been investigated
between nonferromagnetic electrodes.

Ferromagnetic semiconductors, in particular the Eu chalcogenides, have been of
great interest because of their magnetic, optical, and transport properties
\cite{Wachter,Mauger}. Tunneling experiments using these materials as barriers
demonstrated the spin polarization in the tunnel current by observing the
decrease of junction resistance below the Curie temperature ($T_c$) of the
barriers. Esaki {\it et al}. \cite{Esaki} studied the internal field-emission
on metal-EuS-metal junctions at temperatures above and below EuS Curie
temperature which is at 16.5 K for pure annealed thin films. They observed an
increase of field-emission current as the temperature was lowered to below the
magnetic ordering temperature of the barrier and interpreted it as caused by
the decrease of barrier height when spin ordering takes place. Field-emission
studies \cite{Muller,Kisker1,Baum,Kisker2} on EuS-coated tungsten tips showed a high degree of
polarization of the field-emitted electrons below the Curie temperature of EuS.
In these experiments when EuS is used as a tunnel barrier, the conduction
band splits into spin-up and spin-down subbands and the barrier height for
these two subbands is changed. Since the tunneling process depends sensitively
on the barrier height, the splitting of the EuS conduction band greatly increases
the probability of tunneling for spin-up electrons and reduces that for
spin-down electrons. This is called the ``spin-filter'' effect \cite{Moodera}.
In favorable cases \cite{Moodera2} the spin polarization in tunneling has
exceeded 99$\%$.

Electron-Spin Polarization, was studied theoretically by Takahashi
\cite{Takahashi2}. Using the CPA for the {\it s-f} model he presented the
dependence of the ESP on the magnetic field, the temperature, and the energy in
the parabolic band model. Recently Metzke and Nolting \cite{Metzke} presented a
new interpretation of the spin-filter effect using many-body theory on the
{\it s-f} model in the film geometry, including an strong external electric
field.

The purpose of the present paper is to develop a better theoretical
framework for spin-polarized tunneling based on the coherent potential
theory for the {\it s-f} model in the single band tight-binding model. Using
this formalism we investigate density of states and tunneling spin-polarized
through a EuS monolayer and the numerical calculated results are compared with
the results of some experiments.

\section{Model and formalism}
We consider a system consisting of a ferromagnetic semiconductor (EuS) thin
layer sandwiched between two semi-infinite lead wires. Both the thin layer and
lead wires are described by a single-orbital tight-binding model with nearest
neighbor hopping {\it t} on a simple cubic lattice with lattice constant
{\it a}. We choose the (001) axis of the simple cubic structure to be normal to
the layer and this direction is called {\it z}-direction hereafter.

We use the {\it s-f} (or {\it s-d}) model as it is believed to yield a good
description for magnetic semiconductors. In this model the following
Hamiltonian is used to describe the present system:
\begin{equation}
H=H_s+H_{sf}+H_f,
\end{equation}
\begin{equation}
H_s=-t\sum_{{\bf r}n,{\bf r}\p n\p,\s}c^{\dag}_{{\bf r},n,\s}c_{{\bf r}\p,n\p,
\s}\uu~,
\end{equation}
\begin{equation}
H_{sf}=-I\sum_{{\bf r},\s,\s\p}
({\bf\mbox{\large$\s$}\cdot S}_{{\bf r},0})_{\s\s\p}\uu c^{\dag}_{{\bf r},0,
\s}c_{{\bf r},0,\s\p}\uu~,
\end{equation}
\begin{equation}
H_f=-\sum_{{\bf r},{\bf r}\p}J_{{\bf r}0,{\bf r}\p 0}{\bf S}_{{\bf r},0}
\cdot{\bf S}_{{\bf r}\p,0}-g{\mu_{_B}}{H_0}\sum_{\bf r}S^{z}_{{\bf r},0}~,
\end{equation}
where $\bf r$ and $n$ denote the position in $x$-$y$ plane and the layer index
in the $z$-direction, respectively. Here $H_s$ is the transfer energy
of an $s$-electron between nearest-neighbor sites, $c^{\dag}_{{\bf r},n,\s}$
$(c_{{\bf r},n,\s}^{\uu})$ is a creation (an annihilation) operator of an
$s$-electron with spin $\s (=\up ,\dd)$ at site $({\bf r},n)$; $H_{sf}$ is the
{\it s-f} exchange interaction between the $s$-electron and the $f$-spin where
$\large\s$ is a conduction electron spin operator, and $\it I$ is the {\it s-f}
exchange interaction energy. Each lattice point of the film is occupied by a
localized magnetic moment, represented by a spin operator ${\bf S}_{{\bf r},0}$.
The first term in (4) describes the direct exchange coupling of the Heisenberg
type between these localized moments where $J_{{\bf r}0,{\bf r}\p 0}$ is an
exchange integral; and the second term is the Zeeman energy when a magnetic
field is applied in the {\it z}-direction. {\it g} is the Land{\'e} factor and
$\mu_{_B}$ is a Bohr magneton. The Zeeman effect on $s$-electrons are completely
ignored. In this study $H_f$ is treated in the molecular field approximation
to obtain the magnetization at each temperature, ignoring the $f$-spin
correlation. Thus we can define a normalized magnetic field in terms of the applied magnetic
field by
\begin{equation}
h=\frac{(S+1)g\mu_{_{B}}H_0}{3k_{_{B}}T_{c}}.
\end{equation}
We must note that $h$=0.06 corresponds to the magnetic field of 5 kOe.

The investigation of the spin-disorder scattering on the conduction electron
in the bulk europium chalcogenides within the single-site CPA is not new, and
was first performed by Nolting \cite{Nolt}. In this approximation in
order to treat electron scattering by thermally fluctuating localized spin
at the layer, we consider a single $f$-spin located at
site {\bf r} in an effective layered medium described by an effective potential
(or coherent potential) which is site diagonal and takes the value $\ss_{\up}$
or $\ss_{\dd}$, according to the spin orientation of the $\it s$-electron. As
in \cite{Takahashi1}, we apply the condition that the average scattering of the
$s$-electron by thermally fluctuating $f$-spin in the medium is zero.
Thus we define the {\it t}-matrix of the {\it s-f} exchange interaction as
\begin{equation}
t_{\bf r}=v_{\bf r}(1-{\bar G}v_{\bf r})^{-1}  ,
\end{equation}
where $\bar G$ is the effective Green's function of the layer in question.
Here $t_{\bf r}$ is the complete scattering associated with the isolated
potential $v_{\bf r}$ in the effective medium which is expressed as
\begin{equation}
v_{\bf r}=\sum_{\s\s\p}[-I({\bf\mbox{\large$\s$}\cdot S}_{\bf r})_{\s\s\p}\uu
-\ss_{\s}\delta_{\s\s\p}]c^{\dag}_{{\bf r},0,\s}c_{{\bf r},0,\s\p}\uu.
\end{equation}

We assume that the system is large enough and has translational invariance
along {\it x, y} directions. We impose periodic boundary conditions along these
directions and use the $|{\bf k_\parallel},n\ras$ representation, where
${\bf k_\parallel}=(k_{x},k_{y})$ is {\it x-y} components of the electron
wave vector. In this representation the effective Green's function is
{\bf k}-diagonal and can be given by \cite{Brataas}
\begin{eqnarray}
F_\s(Z_+)&=&\frac{1}{N_\parallel}\sum_{\bf k_\parallel}\las{\bf k_\parallel},0,
\s|{\bar G}|{\bf k_\parallel},0,\s\ras \nonumber\\
&=&\frac{1}{N_\parallel}\sum_{\bf k_\parallel}\frac{1}
{[G^{+(0)}_{\bf k_\parallel}]^{-1}-\ss_\s}  ,
\label{gf}
\end{eqnarray}
where $N_\parallel$ is the number of sites in the plane of the layer,
$Z_{\pm}=E \pm i\delta$ and $G^{+(0)}_{\bf k_\parallel}$ is the unperturbed
Green's function propagating from left $(n<0)$ to right $(n>0)$ and is
expressed as
\cite{Fisher,Itoh1}
\begin{equation}
G^{+(0)}_{\bf k_\parallel}=\frac{1}{i(2ta)\si k_{\perp}a},
\end{equation}
where
\begin{equation}
-2t\co k_{\perp}a=E+2t(\co k_xa+\co k_ya).
\end{equation}
Here, $\delta$ is a small positive number and the summation includes both
propagating and evanescent states.

Using the elements of the {\it t}-matrix of the {\it s-f} exchange interaction,
the coherent potential, $\ss_\s$, in the thin layer (at {\it n}=0) can
be determined by the following equations \cite{Takahashi1}:
\begin{equation}
{\Big\las}\las{\bf k_\parallel},0,\up|t_{\bf r}|{\bf k_\parallel},0,\up
\ras{\Big\ras}_{T}\uu=0  ,
\end{equation}
\begin{equation}
{\Big\las}\las{\bf k_\parallel},0,\dd|t_{\bf r}|{\bf k_\parallel},0,\dd
\ras{\Big\ras}_{T}\uu=0  ,
\end{equation}
where the bracket $\las\cdots\ras_{T}\uu$ means the thermal average.
These equations can be transformed into the equations to determine $\ss_\up$
and $\ss_\dd$ \cite{Takahashi1}. Once $\ss_\s$ is determined, $F_{\s}$ is
obtained using Eq. (\ref{gf}). In the single-site approximation, the tunneling
density of states for spin $\s$ electron is calculated by
\begin{equation}
D_{\s}(E)=-\frac{1}{\pi}~\mbox{Im}~F_{\s}(E +i\delta)  ,
\end{equation}
and should satisfy the following equation in all of the present numerical
calculations
\begin{equation}
\int_{-\infty}^{+\infty} D_{\s}(E)dE=1.0~.
\end{equation}

We must note that in this work the spin flip of the $s$-electron
is taken into account in the $t$-matrix formula, but the spin flip of the
$f$-spin is neglected because the $f$-spin is treated as a classical spin.

We are now at the position to calculate the actual spin polarization for a
ferromagnetic semiconductor as a function of temperature. An ensemble of
electrons is said to have electron spin polarization when they show a
preferential spin direction. ESP is described by the vector ${\bf P}$ in the
preferential direction, whose magnitude is given by
\begin{equation}
P=\frac{N_\up-N_\dd}{N_\up+N_\dd},
\end{equation}
where $N_\up (N_\dd)$ is the number of emitted electrons with spin-up (down).
This quantity is obtained in experiments when the conduction band of EuS is
almost empty. Then it is reasonable to assume that
$N_\up/N_\dd$ is equal to $D_\up(E_F)/D_\dd(E_F)$, where $D_\s(E_F)$ is the
density of states at the Fermi energy. Using these definitions, the polarization
of the tunneling density of states is given by
\begin{equation}
P=\frac{D_{\up}(E_F)-D_{\dd}(E_F)}{D_{\up}(E_F)+D_{\dd}(E_F)}  .
\end{equation}
In this investigation $E_F$ corresponds to the energy equivalent to almost the
bottom of the conduction band.

\section{Numerical results}
In numerical calculations we adopted the following values for EuS: $a$
=5.97 \AA, $T_c$=16.5 K \cite{Baum}, $S$=7/2, $I$=0.1 eV, $W$=0.9 eV \cite
{Nolting}. Here $W$ is bandwidth and is 12$t$. Note that following Refs. \cite
{Brataas} and \cite{Fisher}, the energy is measured in units of $ta$. The small
imaginary part of the energy is chosen $\delta$=0.02 to simplify the numerical
calculations.

In Fig. 1, the results of the density of states calculated in the present
study is depicted as a function of the energy for $T$=0, 0.85 $T_c$ and $T_c$
Fig. 1(a)-(c) with no external magnetic field and at $T$=1.2 $T_c$ Fig. 1(d)
with a magnetic field of 5 kOe. According to the first order perturbation theory
of the EuS conduction band, the ideal spin polarization of 100$\%$ for all
temperatures below $T_c$ and 0$\%$ above is expected where the band splits into
two completely spin-polarized subbands.
Present results show that at $T\ge T_c$ when magnetic field is zero,
magnetization is zero and there is not any spin splitting between the two
subbands.
As the magnetization increases from paramagnetic state, the density of
states for up-spin band shifts upward and for down-spin band shifts downward.
The reason is that by decreasing the temperature the number of up-spin electrons
participate in tunneling procedure increase but for down-spin it decreases.

\begin{figure}
\begin{center}
\leavevmode\hbox{\epsfxsize=0.48\textwidth\epsffile{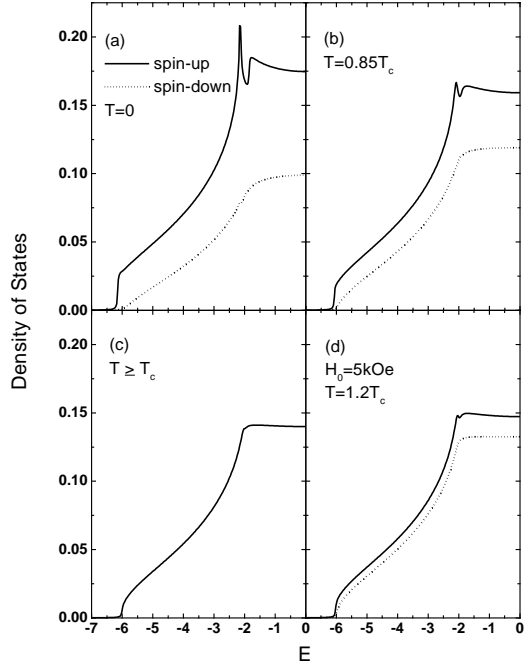}}
\end{center}
\caption{The density of states as a function of energy without (curve $a-c$)
and with the applied magnetic field (curve-$d$) at the various temperatures.
Here only the results for ${\it E_F\le}$ 0 are shown since the results for ${\it E_F\ge}$ 0
are symmetric about the band center $E_F$=0.}
\end{figure}

These curves show that a spin gap dose not exist for both the conduction
electron directions, i.e., even at zero temperature the spin-down band has a tail
extending down to the spin-up band edge as is already evidenced in CPA results
of Nolting \cite{Nolt} in bulk europium chalcogenides. Consequently the relative conduction
electron spin polarization does not reach the ideal value of 100$\%$. The
existence of the band tail suggests the spin-flip scattering to play an
important role in the spin-polarized transport. The oscillations at the density
of states are the effects of the van Hove singularities which appear
at $E$=$\pm$2.0 .

In this investigation the spin flip of the $f$-spin was ignored since the
$f$-spin was treated as a classical spin. Thus the present study cannot
recover the existence of a spin polaron peak in the density of states as Metzke
and Nolting \cite{Metzke} reported.

When the temperature is far below $T_c$, the spin filter effect in EuS
polarizes the tunnel current even when no magnetic field is applied. The effect
of the external magnetic field is only to remove the domain structure.
In Fig. 2(a) and (b), we show the results for tunneling spin polarized for $E_F$=-6.07 as
a function of temperature together with the field-ESP experiment data for EuS
at $H_0$=0 \cite{Kisker1,Baum} and $H_0$=5 kOe \cite{Kisker2} respectively. The
agreement between the calculated results and experiment data is satisfactory.
The difference between experiment results and present theory at around $T_c$
and at paramagnetic temperatures is due to ignoring the exchange scattering
due to the correlation of $f$-spins and use of monolayer instead of multilayers.
The inset of Fig. 2(b) shows the tunneling spin polarized as a function of
$s$-electron energy at $T$=0. This result, together with the experiment data,
strongly suggest that the ESP in field-emission studies corresponds to the $P$
value for low electron energy ${\it (E\le}$ -5.8 in this study). For energy less
than this value the ESP is as high as 90$\%$ even at zero magnetic field.

On the other hand, experimental results show that the degree of field-ESP
observed is much higher than that of photoemission (photo-ESP)
\cite{Wachter,Mauger}. In photo-ESP the electrons from $4f^7$ states by
absorbing a photon transfer to the conduction band. Thus, most photoemitted
electrons have higher energy than the energy of the bottom of the conduction
band and are related to a large density of states. Therefore, as Fig. 1 and
the inset of Fig. 2(b) show the degree of polarization of high-energy electrons
such as -5.8$\le E \le$-5.0, in the present study, is rather low.

Consequently the difference between the results of photo-ESP and field-ESP is
due to the difference in the energies of electrons in the emission processes,
which is consistent with the work of Takahashi \cite{Takahashi2} in bulk
europium chalcogenides. These results are confirmed by experimental results of
Kisker {\it et al}. \cite{Kisker2} in which they showed that the low energy electrons
are highly polarized, whereas the polarization degree of high energy electrons
is rather low.

\begin{figure}
\begin{center}
\leavevmode\hbox{\epsfxsize=0.48\textwidth\epsffile{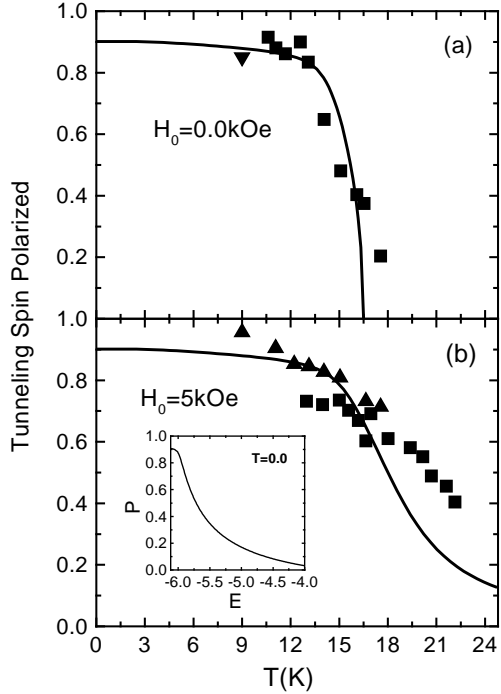}}
\end{center}
\caption{The tunneling spin polarized as a function of temperature without
(curve-$a$) and with the applied magnetic field (curve-$b$). The experimental
results in curve-$a$ (square and down-triangle symbols) are taken from Ref. 8
and Ref. 9 respectively and in curve-$b$ (square and up-triangle) are taken from
Ref. 10. The inset is a calculated result of the ESP as a function of energy.}
\end{figure}

\section{Concluding remarks}
In this study we attempted to explain the results of the tunneling spin
polarized measurements based on the spin-polarized subbands picture for an EuS
layer. Using the CPA for the {\it s-f} model we considered the effects of spin
disorder on the perpendicular transport through a monolayer. Assuming a single
band tight-binding model numerical calculations were performed for the density
of states and tunneling spin polarized. The agreement between the calculated
and some experiments results is satisfactory.

Throughout this investigation the effect of interface roughness that plays an
important role in the giant magnetoresistance and the scattering due to the
correlation of $f$-spins are not taken into account.


\begin{thebibliography}{plane}
\bibitem {Baibich}
{M. N. Baibich, J. M. Broto, A. Fert, Nguyen Van Dau, F. Petroff, P. Etienne,
G. Creuzet, A. Friederich, and J. Chazelas, Phys. Rev. Lett. {\bf 61}
(1988) 2472.}

\bibitem{Julliere}
{M. Julliere, Phys. Lett. A {\bf 54} (1975) 225.}

\bibitem{Gr}
{P. Gr\"unberg, R. Schreiber, Y. Pang, M. B. Brodsky, and H. Sowers, Phys. Rev. Lett.
{\bf 57} (1986) 2442.}

\bibitem{Wachter}
{P. Wachter, Handbook on the Physics and Chemistry of Rare Earths,
edited by K. A. Gschneider, Jr. and L. Eyring, North-Holland, Amsterdam,
1979 (Chapter 19).}

\bibitem{Mauger}
{A. Mauger and C. Godart, Phys. Rep. {\bf 141} (1986) 51.}

\bibitem{Esaki}
{L. Esaki, P. J. Stiles, and S. von Molnar, Phys. Rev. Lett. {\bf 19}
(1967) 852.}

\bibitem{Muller}
{N. M\"uller, W. Eckstein, W. Heiland, and W. Zinn, Phys. Rev. Lett.
{\bf 29} (1972) 1651.}

\bibitem{Kisker1}
{E. Kisker, G. Baum, A. H. Mahan, W. Raith, and K. Schr\"oder,
Phys. Rev. Lett. {\bf 36} (1976) 982.}

\bibitem{Baum}
{G. Baum, E. Kisker, A. H. Mahan, W. Raith, and B. Reihl, Appl. Phys.
{\bf 14} (1977) 149.}

\bibitem{Kisker2}
{E. Kisker, G. Baum, A. H. Mahan, W. Raith, and B. Reihl, Phys. Rev. B
{\bf 18} (1978) 2256.}

\bibitem{Moodera}
{J. S. Moodera, X. Hao, G. A. Gibson, and R. Meservey, Phys. Rev. Lett.
{\bf 61} (1988) 637; X. Hao, J. S. Moodera, and R. Meservey, Phys. Rev. B
{\bf 42} (1990) 8235.}

\bibitem{Moodera2}
{J. S. Moodera, R. Meservey, and X. Hao, Phys. Rev. Lett. {\bf 70} (1993) 853.}

\bibitem{Takahashi2}
{M. Takahashi, Phys. Rev. B {\bf 56} (1997) 7389.}


\bibitem{Metzke}
{R. Metzke and W. Nolting, Phys. Rev. B {\bf 58} (1998) 8579.}

\bibitem{Nolt}
{W. Nolting, Phys. Status Solidi B {\bf 96} (1979) 11.}

\bibitem{Takahashi1}
{M. Takahashi and K. Mitsui, Phys. Rev. B {\bf 54} (1996) 11298.}

\bibitem{Brataas}
{A. Brataas and G. E. W. Bauer, Phys. Rev. B {\bf 49} (1994) 14684.}

\bibitem{Fisher}
{D. S. Fisher and P. A. Lee, Phys. Rev. B {\bf 23} (1981) 6851.}

\bibitem{Itoh1}
{H. Itoh, J. Inoue, Y. Asano, and S. Maekawa, J. Magn. Magn. Mater.
{\bf 156} (1996) 343.}

\bibitem{Nolting}
{W. Nolting, U. Dubil, and M. Matlak, J. Phys. C {\bf 18} (1985) 3687.}

\end{thebibliography}
\end{document}